\documentclass[12pt]{article}

\usepackage{amsmath}
\usepackage{amssymb}
\usepackage{bm}
\usepackage{graphicx}
\usepackage{xcolor}
\usepackage{caption}
\usepackage{subcaption}
\usepackage{tikz}
\usetikzlibrary{arrows.meta,calc,positioning}
\usepackage{pgfplots}
\pgfplotsset{compat=1.16}

\usepackage[braket,qm]{qcircuit}

\usepackage[
    a4paper,
    left=2.0cm,
    right=2 cm,
    top=1.5cm,
    bottom=2.0 cm
]{geometry}

\begin{document}

\title{Quantum computation of energy level degeneracy and thermodynamical properties of spin systems}

\author{
V.~M.~Tkachuk\thanks{Corresponding author: \texttt{voltkachuk@gmail.com}},
\quad M.~I.~Samar
\\[0.3em]
\small $^{1}$Department for Theoretical Physics,
Ivan Franko National University of Lviv, Ukraine
\\[-0.2em]
\small $^{2}$R\&D Department, SoftServe, Ukraine
}
\date{}
\maketitle
\begin{abstract}
We propose a hybrid quantum--classical method for reconstructing
energy-level degeneracies and thermodynamic properties of spin systems.
A single ancilla qubit is used to measure the partition function
at imaginary inverse temperatures. For spectra lying on a known
equally spaced energy grid, the inverse discrete Fourier transform
allows exact reconstruction of degeneracies in the absence of noise,
without spectral leakage. The quantum circuits are optimized
using commuting interactions and Eulerian ordering.
We demonstrate the method for frustrated Ising antiferromagnets
on periodic Kagome graphs with up to 36 spins, using the
\texttt{ibm\_kingston} quantum processor.
Despite substantial circuit depth and hardware noise,
the reconstructed spectra retain their dominant features,
and the calculated heat capacities reproduce the main
temperature dependence of the exact results.

\par\vspace{0.5em}
\par\vspace{0.5em}

\end{abstract}

\section{Introduction}

The energy-level degeneracies of a spin system provide a direct
route to its equilibrium thermodynamics. Once the energies $E_k$
and their degeneracies $g_k$ are known, the partition function
\begin{equation}
Z(\beta)=\sum_k g_k e^{-\beta E_k},
\qquad \beta=(k_B T)^{-1},
\end{equation}
determines the free energy, internal energy, and heat capacity.
Reconstructing the density of states is therefore useful because
the same spectral information can be reused at different
temperatures. Classical methods, including Wang--Landau sampling
and polynomial expansions, provide effective approaches to this
task~\cite{wang2001,weisse2006}. In particular, Yu applied
Wang--Landau sampling to frustrated Ising antiferromagnets on
Archimedean lattices, including the kagome lattice, obtaining
their densities of states, heat capacities, and residual
entropies~\cite{yu2015}.

Quantum approaches provide alternative ways to access partition
functions and spectral information. Mappings between classical
spin models and quantum-state overlaps establish a connection
between statistical mechanics and quantum
computation~\cite{vandenNest2007}, which has been used to
construct algorithms for additive approximations of Ising
partition functions~\cite{matsuo2014}. Other approaches employ
trace estimation in the one-clean-qubit model~\cite{Chow21}
or encode Ising Boltzmann factors through unitary evolution
in an enlarged system~\cite{laba2023}. Quantum protocols have
also connected partition-function zeros to measurable
dynamical quantities~\cite{Fran21,laba2025,kuzmak2025}.
Complementary advances in analog quantum simulation have
enabled experimental studies of frustrated spin systems.
For example, Mkhitaryan \textit{et al.} investigated the
Shastry--Sutherland Ising model using programmable Rydberg-atom
arrays, observing signatures of quantum-fluctuation-induced
phases~\cite{mkhitaryan2026}.

A closely related direction focuses on reconstructing the
density of states from measurements performed on a quantum
processor. Summer \textit{et al.} developed an approach based
on Chebyshev moments and demonstrated it for a nonintegrable
spin Hamiltonian~\cite{summer2024}. Alternatively,
Ferris \textit{et al.} proposed spectral estimation from the
real-time evolution of a maximally mixed state~\cite{ferris2023},
and Van Goffrier \textit{et al.} applied this approach to an
interacting lattice gauge theory~\cite{vangoffrier2026}.
Related methods extract energy levels from the evolution
of a probe spin or a suitably chosen
observable~\cite{gnatenko_probe,gnatenko_quantity}.

For a general spectrum, Fourier reconstruction from a finite
time interval introduces a finite energy resolution and
spectral broadening. Additional structure becomes available
when all energies lie on a known, equally spaced grid.
If the spacing and a bandwidth containing the entire spectrum
are known, the sampling times can be chosen so that the
spectral weights are recovered by an exact finite Fourier
inversion in the absence of noise.
The relevant distinction is the use of known spectral support
to select the sampling scheme: a discrete Fourier transform
(DFT) can also be used numerically in general spectral-estimation
methods. Matching the sampling times to the energy grid removes
finite-window broadening and spectral leakage under the stated
assumptions, while statistical and hardware errors remain.

This spectral structure arises naturally in Ising models with
commensurate interaction strengths. When the couplings are
bounded integer multiples of a common energy scale, the number
of candidate energy levels grows at most linearly with the
number of interaction terms, whereas the number of spin
configurations grows exponentially with system size.
The reconstruction thus reduces to determining a finite set
of energy-level degeneracies. Knowing the candidate energy
grid, however, does not determine which positions are occupied
or their degeneracies. These weights contain the many-body
counting information required for thermodynamics, and their
reconstruction remains sensitive to measurement precision,
particularly for weak spectral contributions.

The kagome Ising antiferromagnet provides a useful setting
for examining this distinction. It combines a commensurate
spectrum with geometrical frustration and extensive
ground-state degeneracy~\cite{kano1953}. Its exact finite-cluster
spectra provide benchmarks for assessing quantum reconstruction.
Moreover, errors in the relative weights of low-lying levels
can strongly affect low-temperature thermodynamics, making
the heat capacity a useful test of the information retained
in the reconstructed spectrum.

\subsection{Main contributions}

In this work, we combine single-ancilla measurements of the
partition function at imaginary inverse temperatures with
finite Fourier reconstruction and apply the resulting protocol
to frustrated Ising systems. Our main contributions are
the following.

\begin{enumerate}

\item \textbf{Reconstruction matched to a known energy grid.}
We specify a finite set of sampling times for recovering
energy-level degeneracies by the inverse DFT.
The reconstruction is exact for noiseless measurements,
provided that the known grid contains the entire spectrum
without aliasing. For the periodic kagome clusters considered
here, the implemented scheme uses $Q=N+1$ sampling times
for $N$ spins. Since the Hamiltonians are diagonal in the
computational basis, an equal-superposition product state
of the register suffices to measure the normalized trace
through a single ancilla.

\item \textbf{Circuit simplification for commuting interactions.}
We exploit the commutativity of the Ising terms to order
the bond interactions along an Euler circuit and cancel
adjacent parity gates. For the interaction graphs considered,
this reduces the logical controlled-NOT count from $4L$
to $2L+2$, where $L$ is the number of bonds.
The construction implements the evolution without Trotter
error, although hardware routing requires additional gates.

\item \textbf{Experimental reconstruction of spectra and
thermodynamics.}
We demonstrate the protocol on a superconducting quantum
processor for a five-spin benchmark and periodic kagome
clusters containing $6$, $12$, $24$, and $36$ spins.
The reconstructed spectral weights are used to calculate
the heat capacity over a range of temperatures and are
compared with exact finite-cluster results.
The experiments recover the dominant spectral contributions
and the broad heat-capacity structure, with increasing
quantitative deviations for the larger clusters.

\item \textbf{Analysis of measurement precision and weak
spectral weights.}
For independent measurements with $M$ shots per ancilla
component at each of the $Q$ sampling times, we derive
the bound $\sigma_{w_k}\leq 1/\sqrt{QM}$ for the unprocessed
linear estimator of $w_k=g_k/2^N$.
This distinguishes the number of sampling times from
the precision required to resolve individual weights.
For the $36$-spin cluster, the reconstruction additionally
uses the known ground-state energy, without imposing
its degeneracy.

\end{enumerate}

Together, these results demonstrate how known spectral structure
can be exploited to reconstruct energy-level degeneracies and
thermodynamic properties on a quantum processor. The combination
of exact finite Fourier inversion, circuit simplification,
and experimental validation on frustrated Ising clusters
provides a practical framework for this task, with explicit
measurement-precision estimates guiding the resolution
of individual spectral contributions.

The paper is organized as follows. Section~II describes
the quantum protocol for measuring the partition function
at imaginary inverse temperatures. Section~III presents
the reconstruction of energy-level degeneracies using
the inverse DFT and the calculation of thermodynamic
properties. Section~IV demonstrates the method on a
five-spin benchmark, while Sec.~V presents the results
for periodic kagome Ising clusters of different sizes.
Finally, Sec.~VI discusses the advantages, limitations,
and possible extensions of the approach.

\section{Partition Function with Imaginary Temperature and Ancilla-Qubit Representation}

Consider a system of \(N\) spin-\(1/2\) particles described by the
Ising-type Hamiltonian
\begin{eqnarray}
\hat H_z = H(\sigma^z_1,\sigma^z_2,\ldots,\sigma^z_N),
\end{eqnarray}
where \(H\) is a function of the \(z\)-components of the spin operators.

The partition function at imaginary inverse temperature $\beta=it$ reads
\begin{eqnarray}\label{ZI2}
Z(it)={\rm Tr}_{\sigma}e^{-it\hat H_z}
=
\sum_{\bf x}\langle {\bf x}|e^{-it\hat H_z}|{\bf x}\rangle,
\end{eqnarray}
where $t$ is the actual time of evolution. Here and below we use units
in which \(\hbar=1\).

We introduce the notation
$|{\bf x}\rangle=|x_1,x_2,\ldots,x_N\rangle$ with $x_j=0,1$
for the computational-basis states, corresponding to the spin states
\(\uparrow\) and \(\downarrow\), respectively. These states are eigenstates
of the Hamiltonian, with eigenvalues
$H=H(\sigma_1,\sigma_2,\ldots,\sigma_N)$, where
$\sigma_j^z|x_j\rangle=\sigma_j|x_j\rangle$, with $\sigma_j=1$ for
$x_j=0$ and $\sigma_j=-1$ for $x_j=1$.

Tracing over the eigenstates of the Hamiltonian gives the classical
expression for the partition function at imaginary inverse temperature,
\begin{eqnarray}
Z=\sum_{\sigma_1=\pm 1}\sum_{\sigma_2=\pm 1}
\cdots\sum_{\sigma_N=\pm 1}e^{-itH},
\end{eqnarray}
where the Hamiltonian
$H=H(\sigma_1,\sigma_2,\ldots,\sigma_N)$ is a function of the discrete
variables $\sigma_1,\sigma_2,\ldots,\sigma_N$, each of which takes one
of the two values $\pm1$.

Note that this summation contains $2^N$ terms. Thus, a direct brute-force
evaluation of the sum requires a number of terms that grows exponentially
with \(N\), although particular Hamiltonians may admit more efficient
classical methods.

A quantum computer allows the normalized partition function to be
encoded into expectation values of a single ancilla qubit, thereby
avoiding an explicit summation over all $2^N$ spin configurations.
To implement this idea, we couple an ancilla qubit to the system and
consider the total Hamiltonian \cite{Fran21}
\begin{eqnarray}
\hat H_T=\sigma_0^z\hat H_z
=\sigma_0^z H(\sigma_1^z,\sigma_2^z,\ldots,\sigma_N^z).
\end{eqnarray}

Let us consider the evolution of a quantum state governed by this
Hamiltonian starting from the initial state
$|+\rangle|{+,+,\ldots,+}\rangle$, where $|+\rangle$ is the ancilla
state and $|{+,+,\ldots,+}\rangle$ is the state of $N$ spins.
At time $t/2$, the quantum state reads
\begin{eqnarray}
|\psi(t/2)\rangle
=
e^{-i(t/2)\hat H_T}|+\rangle|{+,+,\ldots,+}\rangle.
\end{eqnarray}

For the expectation value of the $x$ component of the ancilla Pauli
operator in this state, we find
\begin{eqnarray}
\langle\sigma_0^x\rangle=
\langle {+,+,\ldots,+}|\langle +|
e^{i(t/2)\hat H_T}
\sigma_0^x
e^{-i(t/2)\hat H_T}
|+\rangle|{+,+,\ldots,+}\rangle.
\end{eqnarray}
Note that $\sigma_0^x$ anticommutes with $\hat H_T$,
$\{\sigma_0^x,\hat H_T\}=0$. Therefore,
$e^{i(t/2)\hat H_T}\sigma_0^x
=\sigma_0^x e^{-i(t/2)\hat H_T}$, and the expectation value reads
\begin{eqnarray}\label{sigmaz}
\langle\sigma_0^x\rangle 
&=&
\langle {+,+,\ldots,+}|\langle +|
e^{it\sigma_0^z\hat H_z}\sigma_0^x
|+\rangle|{+,+,\ldots,+}\rangle
\nonumber\\
&=&
\langle {+,+,\ldots,+}|\langle +|
e^{it\sigma_0^z\hat H_z}
|+\rangle|{+,+,\ldots,+}\rangle,
\end{eqnarray}
where $\sigma_0^x|+\rangle=|+\rangle$.

Taking into account
\begin{eqnarray}
e^{it\sigma_0^z\hat H_z}
=
\cos(t\hat H_z)+i\sigma_0^z\sin(t\hat H_z)
\end{eqnarray}
and $\langle+|\sigma_0^z|+\rangle=0$, we find
\begin{eqnarray}
\langle\sigma_0^x\rangle
&=&
\langle {+,+,\ldots,+}|
\cos(t\hat H_z)
|{+,+,\ldots,+}\rangle
\nonumber\\
&=&
{1\over 2^N}\operatorname{Re} Z.
\end{eqnarray}

Similarly, since $\sigma_0^y$ also anticommutes with $\hat H_T$,
\begin{eqnarray}
\langle\sigma_0^y\rangle
=
\langle {+,+,\ldots,+}|\langle +|
e^{it\sigma_0^z\hat H_z}\sigma_0^y
|+\rangle|{+,+,\ldots,+}\rangle .
\end{eqnarray}
Using $i\sigma_0^z\sigma_0^y=\sigma_0^x$, together with
$\langle+|\sigma_0^y|+\rangle=0$ and
$\langle+|\sigma_0^x|+\rangle=1$, we obtain
\begin{eqnarray}\label{sigmay}
\langle\sigma_0^y\rangle
=
\langle {+,+,\ldots,+}|
\sin(t\hat H_z)
|{+,+,\ldots,+}\rangle
=
-{1\over 2^N}\operatorname{Im} Z.
\end{eqnarray}

Note that $\sigma_0^z$ commutes with $\hat H_T$. As a result, its
expectation value is conserved during the evolution. Since
$\langle+|\sigma_0^z|+\rangle=0$, we find
\begin{eqnarray}
\langle\sigma_0^z\rangle=0.
\end{eqnarray}

Thus, the partition function is determined by the expectation values
of the ancilla Pauli operators:
\begin{eqnarray}\label{ZmeanAnc}
{1\over 2^N}Z
=
\langle\sigma_0^x\rangle
-i\langle\sigma_0^y\rangle.
\end{eqnarray}
Consequently,
\begin{eqnarray}\label{Z2meanAnc}
{1\over 2^{2N}}|Z|^2
=
\langle\sigma_0^x\rangle^2
+\langle\sigma_0^y\rangle^2
=
\langle {\bm \sigma}_0\rangle^2,
\end{eqnarray}
where we have used $\langle\sigma_0^z\rangle=0$.
These formulas relate the partition function to the Bloch vector of
the ancilla qubit (see, for instance, \cite{Chow21}).

For the pure state of the combined ancilla--system, the entanglement
between the ancilla qubit and the remaining $N$ qubits is completely
determined by the length of the ancilla Bloch vector. Note that according
to Eq.~(\ref{ZmeanAnc}), both the real and imaginary parts of the
partition function can be calculated. For instance, the geometric
measure of entanglement reads \cite{Fry17}
\begin{eqnarray}
E={1\over 2}(1-|\langle {\bm \sigma}_0\rangle|),
\end{eqnarray}
and thus the entanglement and the partition function with imaginary
inverse temperature are related as follows:
\begin{eqnarray}\label{EZ}
E={1\over 2}\left(1-{1\over 2^N}|Z(it)|\right).
\end{eqnarray}

Fisher zeros of the partition function correspond to the case when
$|Z|=0$. From Eq.~(\ref{EZ}), we see that the entanglement takes its
maximal value $E={1\over 2}$ when the partition function is zero,
i.e., at a Fisher zero lying on the imaginary inverse-temperature axis.

The relation between the partition function and the expectation values
of the ancilla Pauli operators, Eq.~(\ref{ZmeanAnc}), therefore allows
one to compute the imaginary-inverse-temperature partition function on
a quantum computer. In particular, measurements of
$\langle\sigma_0^x\rangle$ and $\langle\sigma_0^y\rangle$ provide
direct access to both the real and imaginary parts of $Z(it)$.
As shown below, this information can subsequently be used to reconstruct
the energy-level degeneracies of the system.

\section{Reconstruction of energy-level degeneracies and thermodynamic properties}

We consider Ising-type Hamiltonians whose spectra lie on a known
equally spaced energy grid,
\begin{equation}
\label{E1}
E_n=E_{\mathrm c}+n\Delta E,
\qquad n=-K,\ldots,K,
\end{equation}
where $\Delta E>0$ is the energy spacing. The grid contains
$Q=2K+1$ possible energies, with zero degeneracy assigned
to levels absent from the physical spectrum.

The partition function at imaginary inverse temperature reads
\begin{equation}
Z(it)=\sum_{n=-K}^{K}g_n e^{-iE_nt},
\end{equation}
where $g_n$ denotes the degeneracy of $E_n$.
To reconstruct these degeneracies, we choose the evolution times
\begin{equation}
\label{tm}
t_m=\frac{2\pi m}{Q\Delta E},
\qquad m=-K,\ldots,K.
\end{equation}
At these times, the discrete Fourier orthogonality relation
\begin{equation}
\label{delta}
\frac{1}{Q}\sum_{m=-K}^{K}
e^{i(E_k-E_n)t_m}=\delta_{kn}
\end{equation}
allows us to recover the degeneracies by the inverse DFT,
\begin{equation}
\label{g_k}
g_k=\frac{1}{Q}\sum_{m=-K}^{K}
e^{iE_kt_m}Z(it_m).
\end{equation}
Unlike finite-time approximations to the continuous Fourier
transform, this procedure provides exact reconstruction
from only $Q$ time points in the absence of measurement
noise. The known energy grid eliminates spectral leakage
and the need for long evolution times to improve
energy resolution.

The required partition-function values are obtained from
the ancilla-qubit measurements described in the previous
section:
\begin{equation}
\label{ancilla_relation}
\frac{Z(it_m)}{2^N}
=
\langle\sigma_0^x(t_m)\rangle
-i\langle\sigma_0^y(t_m)\rangle.
\end{equation}
Substitution into Eq.~\eqref{g_k} gives the normalized
spectral weights
\begin{equation}
\label{w_k_final}
w_k=\frac{g_k}{2^N}
=
\frac{1}{Q}\sum_{m=-K}^{K}
e^{iE_kt_m}
\left[
\langle\sigma_0^x(t_m)\rangle
-i\langle\sigma_0^y(t_m)\rangle
\right].
\end{equation}

For $M$ shots per ancilla Pauli component at each of
the $Q$ time points, the total measurement budget is
$S=2QM$. Assuming independent measurements and taking
the real part of the inverse DFT, the statistical
uncertainty is bounded by
\begin{equation}
\label{error_w}
\delta w_k\leq\frac{1}{\sqrt{QM}}
=\sqrt{\frac{2}{S}}.
\end{equation}
For typical weights $w_k=O(1/Q)$, the relative uncertainty
scales as $\delta w_k/w_k=O(\sqrt{Q/M})$.
Maintaining a fixed relative accuracy therefore requires
$M\propto Q$, giving $S\propto Q^2$.
Levels with particularly small degeneracies are more
difficult to resolve because their normalized weights
may be much smaller than $1/Q$.

Finally, the Hermiticity of the Hamiltonian implies
\begin{equation}
Z(-it)=Z(it)^*,
\end{equation}
which leads to
\begin{equation}
\langle\sigma_0^x(-t)\rangle
=\langle\sigma_0^x(t)\rangle,
\qquad
\langle\sigma_0^y(-t)\rangle
=-\langle\sigma_0^y(t)\rangle.
\end{equation}
Although this symmetry allows the negative-time contributions to be inferred from positive-time measurements, reducing the number of time points does not reduce the total shot budget required to achieve the same statistical accuracy, since the shots must be redistributed among the remaining measurements. We therefore retain independent measurements at all $Q=2K+1$ time points, allowing symmetry violations caused by hardware noise to be examined. The measured signals can then be symmetrized before applying the inverse DFT, eliminating spurious imaginary contributions to the reconstructed degeneracies $g_k$.

Once the spectral weights are reconstructed, the
thermodynamic properties can be calculated classically.
At real inverse temperature $\beta=1/(k_BT)$,
the partition function becomes
\begin{equation}
\label{Z_beta_w}
Z(\beta)=2^N z(\beta),
\qquad
z(\beta)=\sum_n w_n e^{-\beta E_n},
\qquad
\sum_n w_n=1.
\end{equation}
Setting $k_B=1$, the free energy is
\begin{equation}
F=-\frac{1}{\beta}
\left[N\ln 2+\ln z(\beta)\right].
\end{equation}
The first two energy moments are
\begin{equation}
\label{energy_moments}
\langle H^r\rangle_\beta
=
\frac{\displaystyle\sum_n E_n^r w_n e^{-\beta E_n}}
{\displaystyle\sum_n w_n e^{-\beta E_n}},
\qquad r=1,2,
\end{equation}
and the heat capacity follows from
\begin{equation}
\label{heat_capacity}
C(\beta)=\beta^2
\left(
\langle H^2\rangle_\beta
-\langle H\rangle_\beta^2
\right).
\end{equation}
Thus, the reconstructed spectral weights provide
the thermodynamic properties over the full temperature
range without additional quantum measurements.

\section{Quantum Circuit Protocol for Extraction of the Complex Partition Function}

\begin{figure}[htbp]
\centering
\begin{tikzpicture}[
    wire/.style={thick, draw=black!80},
    gate/.style={
        draw=black!80,
        fill=black!10,
        rounded corners=2pt,
        minimum size=22pt,
        thick,
        font=\small\bfseries
    },
    biggate/.style={
        draw=black!80,
        fill=black!5,
        rounded corners=3pt,
        thick,
        font=\small\bfseries
    },
    measure/.style={
        draw=black!80,
        fill=black!10,
        rounded corners=2pt,
        minimum size=22pt,
        thick
    }
]

\node at (-1,2) (q0) {$|0\rangle_0$};
\node[gate] (h0) at (0.5,2) {$H$};

\node[biggate, minimum height=80pt, minimum width=80pt]
(u) at (2.6,1)
{$e^{-i\frac{t}{2}\sigma_0^z\hat H_z}$};

\node[gate] (rk) at (4.7,2) {$R_k$};
\node[measure] (m0) at (5.9,2) {};

\draw[thick] (5.7,1.95) arc (180:0:0.2);
\draw[thick,->] (5.9,1.95) -- (6.1,2.2);

\node at (-1,1) (q1) {$|0\rangle_1$};
\node[gate] (h1) at (0.5,1) {$H$};

\node at (-1,0) (qn) {$|0\rangle_N$};
\node[gate] (hn) at (0.5,0) {$H$};

\node at (-1,0.5) {$\vdots$};
\node at (0.5,0.5) {$\vdots$};
\node at (4.7,0.5) {$\vdots$};
\node at (5.9,0.5) {$\vdots$};

\draw[wire] (q0) -- (h0);
\draw[wire] (h0) -- (u.west |- h0.center);
\draw[wire] (u.east |- h0.center) -- (rk);
\draw[wire] (rk) -- (m0);
\draw[thick,double,distance=2pt] (m0) -- (7.0,2);

\draw[wire] (q1) -- (h1);
\draw[wire] (h1) -- (u.west |- h1.center);
\draw[wire] (u.east |- h1.center) -- (7.0,1);

\draw[wire] (qn) -- (hn);
\draw[wire] (hn) -- (u.west |- hn.center);
\draw[wire] (u.east |- hn.center) -- (7.0,0);

\end{tikzpicture}
\caption{Quantum circuit implementing the single-ancilla interferometric
protocol for extracting the normalized complex partition function
$Z(it)/2^N$.}
\label{fig:quantum_circuit_final}
\end{figure}
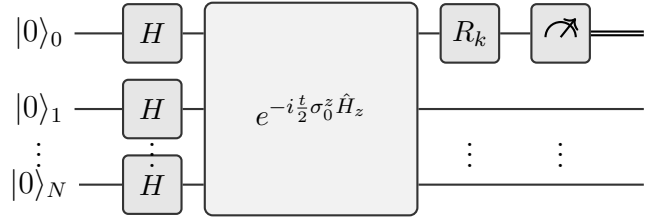

As shown in the preceding sections, the complex partition function
$Z(it)$ can be encoded in the dynamics of a single ancilla qubit,
which acts as an interferometric probe of the many-body system.
This provides direct access to $\operatorname{Tr}e^{-it\hat H}$
without reconstructing the full quantum state. The corresponding
quantum circuit is shown in Fig.~\ref{fig:quantum_circuit_final}.

The circuit consists of one ancilla qubit and $N$ system qubits,
initially prepared in $|0\rangle^{\otimes(N+1)}$. Hadamard gates
prepare the state $|+\rangle^{\otimes(N+1)}$, after which the register
evolves under
\begin{equation}
U_T(t/2)
=
\exp\!\left(
-i\frac{t}{2}\sigma_0^z\hat H_z
\right).
\end{equation}
Finally, the ancilla is rotated to the required measurement basis.
The two settings $R_k=H$ and $R_k=HS^\dagger$ provide
$\langle\sigma_0^x\rangle$ and $\langle\sigma_0^y\rangle$,
respectively, yielding
\begin{equation}
\frac{Z(it)}{2^N}
=
\langle\sigma_0^x(t)\rangle
-i\langle\sigma_0^y(t)\rangle.
\label{eq:ancilla_partition}
\end{equation}

A subsequent inverse discrete Fourier transform of the measured complex
signal reconstructs the spectral weights and therefore the
degeneracies of the energy levels.

\subsection{Five-spin benchmark}

\begin{figure}[htbp]
\centering
\begin{tikzpicture}[
    every node/.style={
        circle,
        draw,
        fill=black,
        inner sep=0pt,
        minimum size=6pt,
        font=\small
    },
    scale=1.5
]

\node[label={below left:4}]  (n4) at (0,0) {};
\node[label={below right:1}] (n1) at (2,0) {};
\node[label={above left:5}]  (n5) at (0,1.5) {};
\node[label={above right:2}] (n2) at (2,1.5) {};
\node[label={above:3}]       (n3) at (1,2.3) {};

\draw[thick] (n1) -- (n2);
\draw[thick] (n1) -- (n4);
\draw[thick] (n1) -- (n5);
\draw[thick] (n2) -- (n3);
\draw[thick] (n2) -- (n4);
\draw[thick] (n2) -- (n5);
\draw[thick] (n3) -- (n5);
\draw[thick] (n4) -- (n5);

\end{tikzpicture}
\caption{Five-vertex ``open-envelope'' graph with
eight Ising interactions.}
\label{fig:ising_graph}
\end{figure}
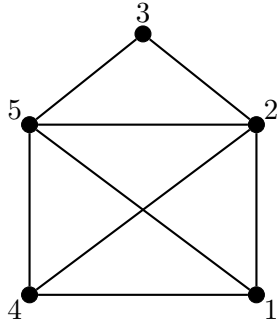

We first benchmark the protocol for the antiferromagnetic
Ising model on the five-vertex ``open-envelope'' graph
shown in Fig.~\ref{fig:ising_graph}. The spins occupy
the vertices, while the $L=8$ links define the pairwise
interactions described by the Hamiltonian
\begin{equation}
\label{Hin}
H=J\sum_{(j,k)\in\mathcal{L}}
\sigma_j^z\sigma_k^z,
\end{equation}
where $\mathcal{L}$ denotes the set of graph edges
and $J>0$ is the antiferromagnetic coupling.

Since the graph contains eight links, the energies
satisfy $-8J\leq E\leq 8J$ and lie on a grid with
spacing $2J$. To test the reconstruction without
assuming the absence of intermediate energy levels,
we use a finer grid,
\begin{equation}
E_n=Jn,\qquad n=-8,\ldots,8,
\end{equation}
corresponding to $E_{\mathrm c}=0$, $\Delta E=J$,
and $K=8$ in Eq.~\eqref{E1}. This choice allows
us to test the recovery of both nonzero degeneracies
and vanishing spectral weights at unoccupied grid points.

According to Eq.~\eqref{tm}, the $Q=17$ sampling
times are
\begin{equation}
\label{t_open_envelope}
t_m=\frac{2\pi m}{17J},
\qquad m=-8,\ldots,8.
\end{equation}

\begin{figure}[htbp]
\centering
\begin{subfigure}[b]{0.49\textwidth}
    \centering
    \includegraphics[width=\textwidth]{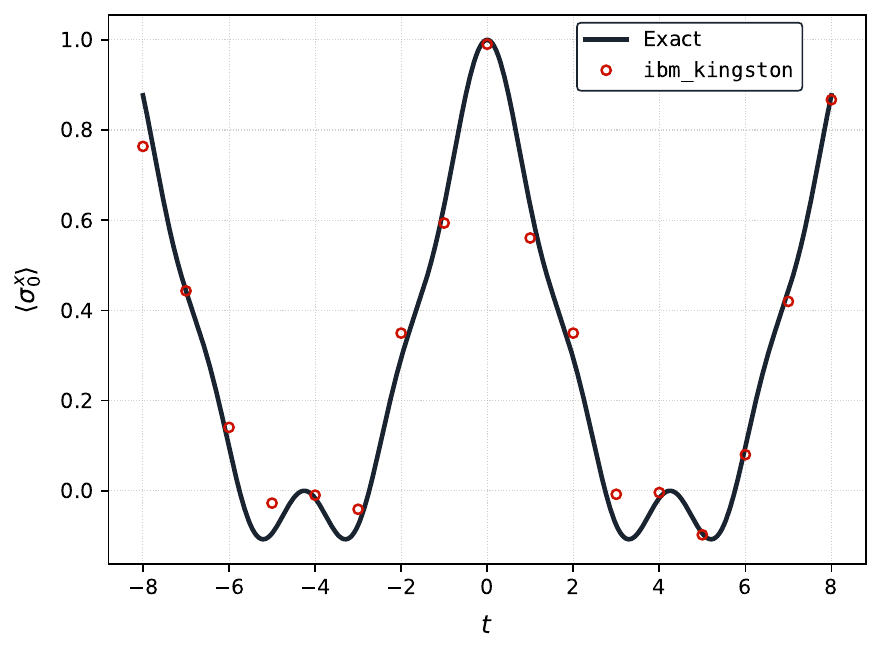}
    \caption{$\langle\sigma_0^x\rangle$.}
    \label{fig:sigma0x_evolution}
\end{subfigure}
\hfill
\begin{subfigure}[b]{0.49\textwidth}
    \centering
    \includegraphics[width=\textwidth]{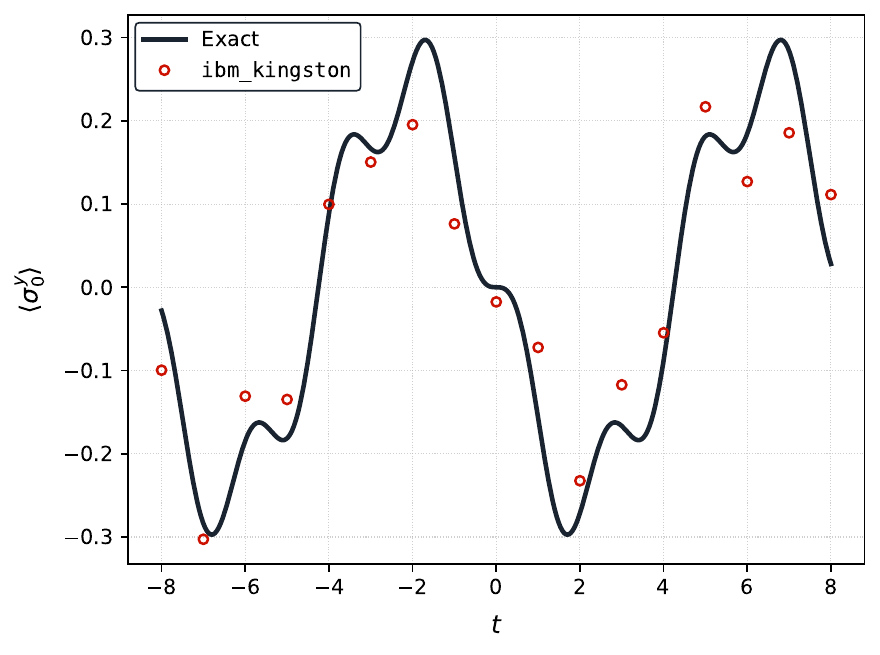}
    \caption{$\langle\sigma_0^y\rangle$.}
    \label{fig:sigma0y_evolution}
\end{subfigure}

\caption{Ancilla expectation values as functions of
evolution time. Exact results are compared with
experimental data from \texttt{ibm\_kingston}.}
\label{fig:sigma_components_comparison}
\end{figure}

The protocol was implemented on the
\texttt{ibm\_kingston} quantum processor with
$1024$ shots per measurement setting.
Exact reference values were calculated from
the energy spectrum obtained by classical
enumeration of all $2^5$ spin configurations.
The corresponding ancilla expectation values
are compared with experimental results in
Fig.~\ref{fig:sigma_components_comparison}.

The discrete Fourier transform yields unconstrained estimates of the
degeneracies. Since positive and negative times are measured
independently, statistical and hardware noise may weakly violate
$Z(-it)=Z(it)^*$ and generate small imaginary parts of the Fourier
coefficients. In the following, only their real parts are used.

Global spin inversion leaves the Hamiltonian invariant,
so all energy levels have even degeneracies satisfying
\begin{equation}
\label{eq:degeneracy_normalization}
g_k\in 2\mathbb{Z}_{\geq0},
\qquad \sum_k g_k=2^N=32.
\end{equation}
To obtain the constrained spectrum, we discard small
Fourier coefficients with $|g_{\rm raw}|<1$, normalize
the remaining values, and constrain them to non-negative
even integers while preserving the total degeneracy.
The resulting values, denoted by $g_{\rm constr}$,
are obtained without prior knowledge of the exact
spectrum, relying only on the symmetry of the Hamiltonian
and the Hilbert-space dimension.

\begin{figure}[htbp]
\centering

\begin{minipage}[b]{0.56\textwidth}
\centering

\begin{tikzpicture}
\begin{axis}[
    xlabel={$E/J$},
    ylabel={Degeneracy},
    xmin=-9.5, xmax=9.5,
    ymin=-1.5, ymax=13,
    xtick={-8,-6,-4,-2,0,2,4,6,8},
    grid=both,
    grid style={line width=.1pt, draw=gray!15},
    major grid style={line width=.2pt, draw=gray!40},
    legend pos=north east,
    legend style={
        font=\tiny,
        nodes={inner sep=0.9pt},
        row sep=0.1ex
    },
    label style={font=\small},
    tick label style={font=\small},
    width=\linewidth,
    height=0.75\linewidth
]

\addplot[
    blue!40!black,
    mark=*,
    mark size=1.8pt,
    ycomb,
    line width=1.8pt
] coordinates {
(-4,2)
(-2,12)
(0,10)
(2,4)
(4,2)
(8,2)
};
\addlegendentry{Exact}

\addplot[
    red,
    mark=o,
    mark size=1.7pt,
    ycomb,
    line width=0.8pt
] coordinates {
(-8,0.427743)
(-7,-0.051487)
(-6,0.488838)
(-5,-0.515886)
(-4,1.593016)
(-3,-0.442884)
(-2,10.571649)
(-1,0.037056)
(0,10.110294)
(1,0.431364)
(2,4.118451)
(3,0.223047)
(4,1.613209)
(5,0.527515)
(6,-0.115076)
(7,0.738155)
(8,1.932496)
};
\addlegendentry{\texttt{ibm\_kingston}}

\end{axis}
\end{tikzpicture}

\par\vspace{2mm}
\makebox[\linewidth][c]{\normalfont (a)}

\end{minipage}%
\hfill
\begin{minipage}[b]{0.40\textwidth}
\centering
\small
\setlength{\tabcolsep}{3.0pt}
\renewcommand{\arraystretch}{1.08}

\resizebox{\linewidth}{!}{%
\begin{tabular}{r|ccc|c}
\hline\hline
$E/J$
&
$g_{\rm raw}$
&
$g_{\rm norm}$
&
$g_{\rm constr}$
&
$g_{\rm exact}$
\\
\hline
$-4$ & 1.593  & 1.703  & 2  & 2  \\
$-2$ & 10.572 & 11.299 & 12 & 12 \\
$0$  & 10.110 & 10.806 & 10 & 10 \\
$2$  & 4.118  & 4.402  & 4  & 4  \\
$4$  & 1.613  & 1.724  & 2  & 2  \\
$8$  & 1.932  & 2.066  & 2  & 2  \\
\hline
\textbf{Total}
& 29.939 & \textbf{32.000}
& \textbf{32} & \textbf{32} \\
\hline\hline
\end{tabular}%
}

\par\vspace{20mm}
\makebox[\linewidth][c]{\normalfont (b)}

\end{minipage}

\vspace{0.25cm}
\caption{
Energy-level degeneracies of the five-spin open-envelope
model.
(a) Raw Fourier reconstruction from
\texttt{ibm\_kingston} compared with the exact spectrum.
(b) Retained coefficients after removing
$|g_{\rm raw}|<1$, their normalized values
$g_{\rm norm}$, constrained degeneracies
$g_{\rm constr}$, and exact values.
}
\label{fig:spectral_reconstruction_5}

\end{figure}

As shown in Fig.~\ref{fig:spectral_reconstruction_5},
the experimental Fourier spectrum exhibits deviations
from the exact degeneracies and small spurious
contributions at unoccupied energy levels.
Nevertheless, the constrained reconstruction
reproduces the exact spectrum,
$g_{\rm constr}=g_{\rm exact}$.
Consequently, the thermodynamic quantities calculated
from these degeneracies also agree exactly with
their theoretical values, and the corresponding curves
are omitted. The effects of imperfect spectral
reconstruction on thermodynamic properties are
examined for larger systems below.

\section{Thermodynamic Reconstruction of the Antiferromagnetic Ising Model on Kagome Lattices}

Having benchmarked the Fourier reconstruction on the five-spin
open-envelope graph, we now apply the protocol to the frustrated
antiferromagnetic Ising model on finite Kagome lattices. In contrast
to the previous benchmark, where the known spacing of the physical
spectrum was deliberately not exploited, here we use the specific
spectral structure of the Kagome model to reduce the number of
required evolution times.

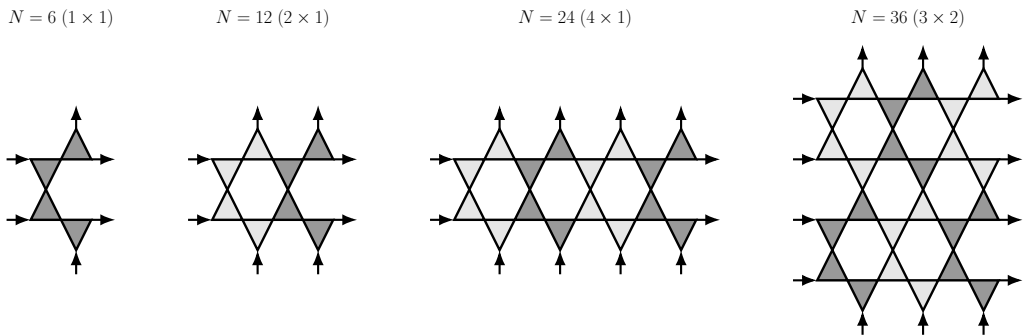
\begin{figure}[htbp]
\centering

\tikzset{
    blockedge/.style={
        black,
        line width=0.9pt
    },
    pbcarrow/.style={
        -{Latex[length=2.0mm]},
        black,
        line width=0.8pt
    }
}

\begin{tikzpicture}[
    scale=0.4,
    every node/.style={transform shape}
]

\newcommand{\kagomeunit}[3]{%
    \begin{scope}[shift={(#1,#2)}]
        \fill[#3] (0,2) -- (1,2) -- (0.5,1) -- cycle;
        \fill[#3] (1.5,3) -- (1,2) -- (2,2) -- cycle;
        \fill[#3] (0.5,1) -- (0,0) -- (1,0) -- cycle;
        \fill[#3] (1.5,-1) -- (1,0) -- (2,0) -- cycle;

        \draw[blockedge] (0,2) -- (1,2) -- (0.5,1) -- cycle;
        \draw[blockedge] (1.5,3) -- (1,2) -- (2,2) -- cycle;
        \draw[blockedge] (0.5,1) -- (0,0) -- (1,0) -- cycle;
        \draw[blockedge] (1.5,-1) -- (1,0) -- (2,0) -- cycle;
    \end{scope}
}

\node[font=\Large, anchor=south] at (1,4.2) {$N=6\;(1\times1)$};

\begin{scope}[shift={(0,-2)}]
\kagomeunit{0}{0}{gray!80}

\draw[pbcarrow] (-0.8,2) -- (0,2);
\draw[pbcarrow] (-0.8,0) -- (0,0);
\draw[pbcarrow] (2,2) -- (2.8,2);
\draw[pbcarrow] (2,0) -- (2.8,0);

\draw[pbcarrow] (1.5,3) -- (1.5,3.8);
\draw[pbcarrow] (1.5,-1.8) -- (1.5,-1);
\end{scope}

\node[font=\Large, anchor=south] at (8,4.2) {$N=12\;(2\times1)$};

\begin{scope}[shift={(0,-2)}]
\kagomeunit{6}{0}{gray!20}
\kagomeunit{8}{0}{gray!80}

\draw[pbcarrow] (5.2,2) -- (6,2);
\draw[pbcarrow] (5.2,0) -- (6,0);
\draw[pbcarrow] (10,2) -- (10.8,2);
\draw[pbcarrow] (10,0) -- (10.8,0);

\draw[pbcarrow] (7.5,3) -- (7.5,3.8);
\draw[pbcarrow] (9.5,3) -- (9.5,3.8);
\draw[pbcarrow] (7.5,-1.8) -- (7.5,-1);
\draw[pbcarrow] (9.5,-1.8) -- (9.5,-1);
\end{scope}

\node[font=\Large, anchor=south] at (18,4.2) {$N=24\;(4\times1)$};

\begin{scope}[shift={(0,-2)}]
\kagomeunit{14}{0}{gray!20}
\kagomeunit{16}{0}{gray!80}
\kagomeunit{18}{0}{gray!20}
\kagomeunit{20}{0}{gray!80}

\draw[pbcarrow] (13.2,2) -- (14,2);
\draw[pbcarrow] (13.2,0) -- (14,0);
\draw[pbcarrow] (22,2) -- (22.8,2);
\draw[pbcarrow] (22,0) -- (22.8,0);

\draw[pbcarrow] (15.5,3) -- (15.5,3.8);
\draw[pbcarrow] (17.5,3) -- (17.5,3.8);
\draw[pbcarrow] (19.5,3) -- (19.5,3.8);
\draw[pbcarrow] (21.5,3) -- (21.5,3.8);

\draw[pbcarrow] (15.5,-1.8) -- (15.5,-1);
\draw[pbcarrow] (17.5,-1.8) -- (17.5,-1);
\draw[pbcarrow] (19.5,-1.8) -- (19.5,-1);
\draw[pbcarrow] (21.5,-1.8) -- (21.5,-1);
\end{scope}

\node[font=\Large, anchor=south] at (29,4.2) {$N=36\;(3\times2)$};

\kagomeunit{26}{0}{gray!20}
\kagomeunit{28}{0}{gray!80}
\kagomeunit{30}{0}{gray!20}

\kagomeunit{26}{-4}{gray!80}
\kagomeunit{28}{-4}{gray!20}
\kagomeunit{30}{-4}{gray!80}

\draw[pbcarrow] (25.2,2) -- (26,2);
\draw[pbcarrow] (25.2,0) -- (26,0);
\draw[pbcarrow] (25.2,-2) -- (26,-2);
\draw[pbcarrow] (25.2,-4) -- (26,-4);

\draw[pbcarrow] (32,2) -- (32.8,2);
\draw[pbcarrow] (32,0) -- (32.8,0);
\draw[pbcarrow] (32,-2) -- (32.8,-2);
\draw[pbcarrow] (32,-4) -- (32.8,-4);

\draw[pbcarrow] (27.5,3) -- (27.5,3.8);
\draw[pbcarrow] (29.5,3) -- (29.5,3.8);
\draw[pbcarrow] (31.5,3) -- (31.5,3.8);

\draw[pbcarrow] (27.5,-5.8) -- (27.5,-5);
\draw[pbcarrow] (29.5,-5.8) -- (29.5,-5);
\draw[pbcarrow] (31.5,-5.8) -- (31.5,-5);

\end{tikzpicture}

\caption{Finite Kagome clusters with $N=6,12,24,$ and $36$ spins,
constructed from six-site building blocks. Periodic boundary
conditions are implemented by identifying the boundary vertices
indicated by the arrows.
The cluster dimensions $L_x\times L_y$ are given in units of
the six-site block.}
\label{fig:kagome_clusters}
\end{figure}

The Kagome antiferromagnet is particularly suitable for this analysis
because geometrical frustration leads to highly degenerate low-energy
levels. Their comparatively large normalized spectral weights improve
the resolution of the corresponding Fourier components in the
presence of finite-shot and hardware noise. We consider periodic
Kagome clusters with $N=6,12,24,$ and $36$ spins, shown in
Fig.~\ref{fig:kagome_clusters}.

\subsection{Kagome spectrum and Fourier grid}

We consider the antiferromagnetic Ising Hamiltonian
\begin{equation}
H
=
J\sum_{\langle i,j\rangle}
\sigma_i^z\sigma_j^z,
\qquad
J>0,
\label{eq:kagome_H}
\end{equation}
where $N$ is the number of spins, $T$ the number of elementary
triangles, and $L$ the number of links. For the
periodic Kagome clusters considered here,
\begin{equation}
N=\frac{3T}{2},
\qquad
L=3T=2N.
\label{eq:kagome_geometry}
\end{equation}

Due to geometrical frustration, each triangle has a minimum energy
$-J$, corresponding to two satisfied links and one frustrated link,
while a triangle with three equal spins contributes $+3J$.
Therefore, if $r$ triangles are in the latter configuration, the
possible energies can be written as
\begin{equation}
E_r
=
J(-T+4r),
\qquad
r=0,1,\ldots,T.
\label{eq:kagome_energy}
\end{equation}
Since $T$ is a multiple of four for the clusters considered here, the
spectrum is contained in the grid
\begin{equation}
E_n=4Jn,
\end{equation}
so that, in the notation of Eq.~\eqref{E1},
\begin{equation}
E_{\mathrm c}=0,
\qquad
\Delta E=4J.
\end{equation}

The physical levels satisfy
\begin{equation}
-\frac{L}{12}\leq n\leq\frac{L}{4}.
\end{equation}
Embedding them into the symmetric Fourier grid $-K\leq n\leq K$
therefore gives
\begin{equation}
K=\frac{L}{4},
\qquad
Q=2K+1=\frac{L}{2}+1.
\label{eq:kagome_K}
\end{equation}

Using Eq.~\eqref{tm}, the sampling times are
\begin{equation}
\label{tau_kagome}
\tau_m
=
\frac{2\pi m}{4J(2K+1)},
\qquad
m=-K,\ldots,K.
\end{equation}
The normalized spectral weights are then reconstructed directly from
Eq.~\eqref{w_k_final}.

In contrast to the five-spin benchmark, here the known spacing
$\Delta E=4J$ gives $K=L/4$, reducing the number of Fourier points
from $2L+1$ to $L/2+1$. An even smaller grid is possible by using
directly the $T+1$ candidate energies of Eq.~\eqref{eq:kagome_energy};
however, we retain the symmetric grid above for the present
calculations.


\subsection{Six-spin Kagome cluster}

We first consider the smallest periodic Kagome cluster with $N=6$.
Exact reference results were calculated from the energy-level
degeneracies obtained by classical enumeration of spin configurations,
while the \texttt{ibm\_kingston} processor was sampled with
$M=2048$ shots per measurement setting.

\begin{figure}[htbp]
\centering

\captionsetup[subfigure]{
    font=footnotesize,
    skip=2pt,
    labelsep=none
}


\begin{subfigure}[b]{0.48\linewidth}
    \centering
    \includegraphics[width=\linewidth]{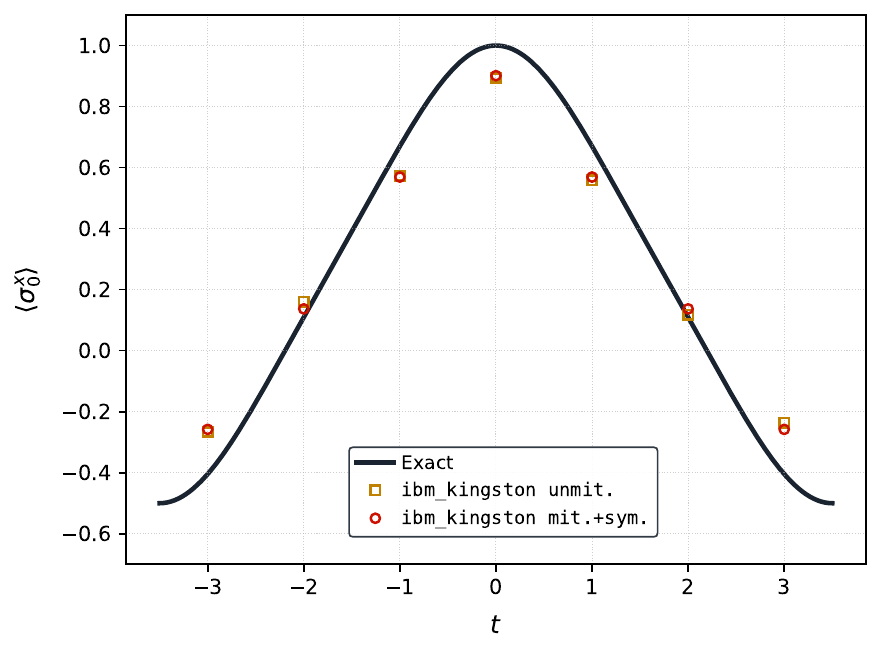}
    \caption{}
    \label{fig:sigma_x_6}
\end{subfigure}%
\hfill%
\begin{subfigure}[b]{0.48\linewidth}
    \centering
    \includegraphics[width=\linewidth]{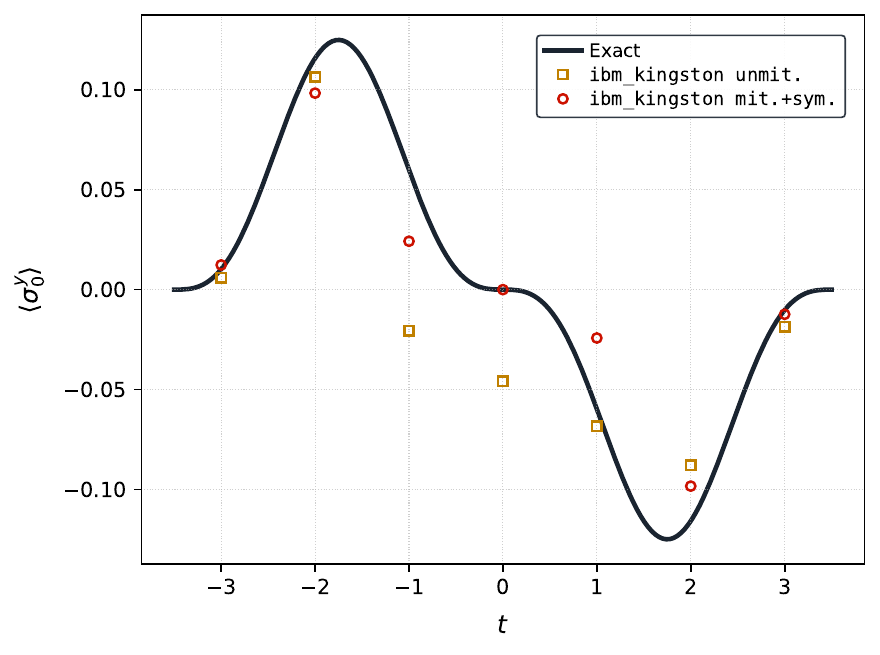}
    \caption{}
    \label{fig:sigma_y_6}
\end{subfigure}

\vspace{0.15cm}


\begin{subfigure}[b]{0.485\linewidth}
\centering

\begin{tikzpicture}
\begin{axis}[
    xlabel={$E/J$},
    ylabel={Degeneracy},
    xmin=-13.5, xmax=13.5,
    ymin=-1.5, ymax=30,
    xtick={-12,-8,-4,0,4,8,12},
    grid=both,
    grid style={line width=.1pt, draw=gray!15},
    major grid style={line width=.2pt, draw=gray!40},
    legend pos=north east,
    legend style={
        font=\tiny,
        nodes={inner sep=0.9pt},
        row sep=0.1ex
    },
    label style={font=\small},
    tick label style={font=\small},
    width=\linewidth,
    height=0.75\linewidth
]

\addplot[
    blue!40!black,
    mark=*,
    mark size=1.8pt,
    ycomb,
    line width=1.8pt
] coordinates {
(-4,26)
(0,16)
(4,20)
(12,2)
};
\addlegendentry{Exact}

\addplot[
    red,
    mark=o,
    mark size=1.7pt,
    ycomb,
    line width=0.8pt
] coordinates {
(-12,0.4826)
(-8,0.2009)
(-4,20.6175)
(0,16.4282)
(4,16.2216)
(8,1.2516)
(12,2.4703)
};
\addlegendentry{\texttt{ibm\_kingston} mit.}

\addplot[
    green!50!black,
    mark=square*,
    mark size=1.7pt,
    ycomb,
    line width=0.8pt
] coordinates {
(-12,0)
(-8,0)
(-4,26)
(0,18)
(4,18)
(8,0)
(12,2)
};
\addlegendentry{\texttt{ibm\_kingston} post.}

\end{axis}
\end{tikzpicture}

\caption{}
\label{fig:spectral_profiles_6}
\end{subfigure}%
\hfill%
\begin{subfigure}[b]{0.485\linewidth}
\centering

\scriptsize
\setlength{\tabcolsep}{1.5pt}
\renewcommand{\arraystretch}{1.02}

\resizebox{\linewidth}{!}{%
\begin{tabular}{r|ccc|ccc|c}
\hline
\hline
&
\multicolumn{6}{c|}{\textbf{\texttt{ibm\_kingston}}}
& \\
\cline{2-7}
&
\multicolumn{3}{c|}{\textbf{Unmitigated}}
&
\multicolumn{3}{c|}{\textbf{Mitigated}}
& \\
\hline

$E/J$
&
$g$
&
$g_{\rm norm}$
&
$g_{\rm post}$
&
$g$
&
$g_{\rm norm}$
&
$g_{\rm post}$
&
$g_{\rm exact}$
\\
\hline

$-12$
& $0.48$ & $0$ & $0$
& $0.48$ & $0$ & $0$
& $0$ \\

$-8$
& $0.20$ & $0$ & $0$
& $0.20$ & $0$ & $0$
& $0$ \\

$-4$
& $20.37$ & $23.58$ & $24$
& $20.62$ & $23.68$ & $26$
& $26$ \\

$0$
& $16.45$ & $19.04$ & $20$
& $16.43$ & $18.86$ & $18$
& $16$ \\

$4$
& $16.02$ & $18.55$ & $18$
& $16.22$ & $18.62$ & $18$
& $20$ \\

$8$
& $1.24$ & $0$ & $0$
& $1.25$ & $0$ & $0$
& $0$ \\

$12$
& $2.44$ & $2.82$ & $2$
& $2.47$ & $2.84$ & $2$
& $2$ \\

\hline
\textbf{Total}
& --
& \textbf{64.00}
& \textbf{64}
& --
& \textbf{64.00}
& \textbf{64}
& \textbf{64}
\\
\hline
\hline
\end{tabular}%
}

\vspace{1.0cm}
\caption{}
\label{fig:spectral_table_6}
\end{subfigure}


\caption{
Results for the six-spin Kagome cluster.
(a) Ancilla expectation value
$\langle\sigma_0^x\rangle$ and
(b) $\langle\sigma_0^y\rangle$
obtained from unmitigated and readout-mitigated,
symmetrized \texttt{ibm\_kingston} data, together with
the exact results.
(c) Degeneracy spectrum reconstructed from the mitigated
data, compared with the post-processed and exact spectra.
(d) Reconstructed degeneracies obtained from the
unmitigated and mitigated data.
The normalized values $g_{\rm norm}$ are obtained
by applying the $3\sigma$ statistical threshold and
rescaling the remaining spectral contributions to satisfy
$\sum_E g(E)=2^N$.
The post-processed values $g_{\rm post}$ additionally satisfy
the imposed discrete constraints.
}
\label{fig:kagome_report_6}

\end{figure}

For the hardware data, we compare the unmitigated ancilla expectation
values with those obtained after readout-error mitigation and
symmetrization.
The corresponding expectation values are shown in
Figs.~\ref{fig:sigma_x_6} and~\ref{fig:sigma_y_6}.

The unmitigated and mitigated ancilla expectation values were
used independently for the Fourier reconstruction of the
degeneracy spectrum. The spectrum obtained from the mitigated
data is shown in Fig.~\ref{fig:spectral_profiles_6},
while the reconstructed degeneracies for both experimental
datasets are summarized in Fig.~\ref{fig:spectral_table_6}.

Finite sampling and hardware noise introduce spurious Fourier
components and deviations from the exact degeneracies.
Furthermore, deviations from the exact time-reversal relations
may generate nonzero imaginary parts in the reconstructed
degeneracies. Since the physical degeneracies are real,
only $\mathrm{Re}(g_k)$ is retained in the subsequent analysis.

The statistical uncertainty associated with finite sampling
can be estimated using the shot-noise bound
\begin{equation}
\sigma_{g_k}
\leq
\frac{2^N}{\sqrt{QM}},
\label{eq:shot_noise_bound}
\end{equation}
where $Q$ is the number of Fourier sampling points and
$M$ is the number of shots per measurement setting.

In this case, $Q=7$ and $M=2048$, yielding
\begin{equation}
\sigma_{g_k}\leq 0.54,
\end{equation}
with a corresponding $3\sigma$ shot-noise threshold
of approximately $1.6$.

To suppress statistically insignificant spectral contributions, we discard coefficients below the $3\sigma$ threshold and renormalize the remaining degeneracies to satisfy $\sum_k g_k=2^N$. This procedure is applied independently to the unmitigated and mitigated datasets, yielding $g_k^{\rm norm}$. The normalized values are then rounded to non-negative even integers, as required by global spin-inversion symmetry. If the total degeneracy differs from $2^N$ after rounding, the difference is added to or subtracted from the largest degeneracy to restore normalization. The resulting values are denoted by $g_k^{\rm post}$.

As shown in Fig.~\ref{fig:spectral_table_6}, both
post-processed spectra reproduce the occupied energy
levels. Readout-error mitigation improves the ground-state
degeneracy, which reaches its exact value after
post-processing, although deviations remain at excited
levels. Thus, physical constraints and readout mitigation
improve the reconstruction but do not completely eliminate
hardware errors.

\begin{figure}[htbp]
\centering
\captionsetup[subfigure]{
    font=footnotesize,
    skip=2pt,
    labelsep=none
}

\begin{subfigure}[b]{0.48\linewidth}
    \centering
    \includegraphics[width=\linewidth]{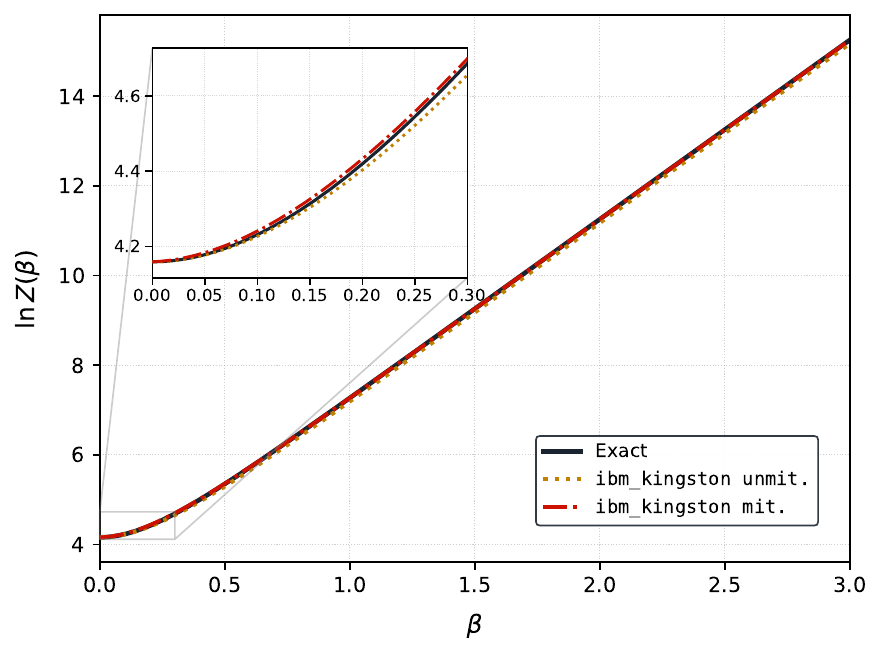}
    \caption{}
    \label{fig:lnZ_6}
\end{subfigure}%
\hfill%
\begin{subfigure}[b]{0.48\linewidth}
    \centering
    \includegraphics[width=\linewidth]{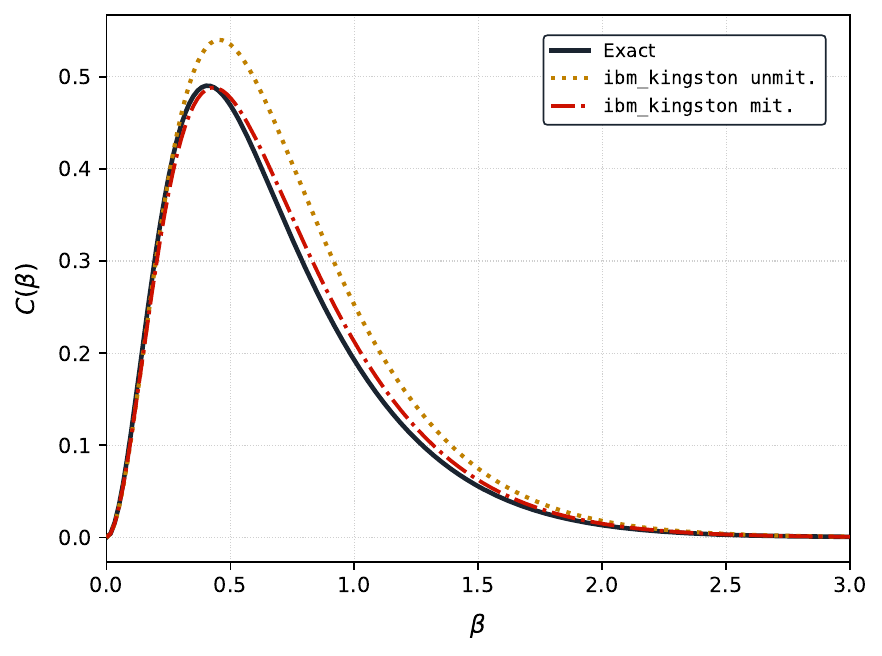}
    \caption{}
    \label{fig:heat_capacity_6}
\end{subfigure}

\caption{
Thermodynamic properties of the six-spin Kagome cluster:
(a) logarithm of the partition function, $\ln Z(\beta)$,
and (b) heat capacity, $C(\beta)$.
Exact results are compared with those reconstructed
from unmitigated and readout-mitigated
\texttt{ibm\_kingston} data after post-processing.
}
\label{fig:thermodynamics_combined_analysis_6}
\end{figure}

The reconstructed degeneracies are subsequently used to
calculate the real-temperature partition function and heat
capacity. Figure~\ref{fig:thermodynamics_combined_analysis_6}
compares the corresponding thermodynamic quantities
obtained from both experimental datasets with the exact
results.

The reconstructed thermodynamic curves remain close to
the exact results. While normalization guarantees the
correct high-temperature limit $Z(0)=2^N$, deviations
at finite temperatures reflect errors in the distribution
of spectral weights, particularly among low-energy levels.
These results demonstrate that thermodynamic properties
can be reliably estimated even when individual
degeneracies are not reconstructed exactly.


\subsection{Twelve-spin Kagome cluster}

We next consider the periodic Kagome cluster with $N=12$.
To assess the influence of hardware noise, we compare
measurements from the \texttt{ibm\_kingston} processor
with shot-based \texttt{AerSimulator} results obtained
without device noise. Both datasets use $M=2048$ shots
per measurement setting, allowing us to distinguish
finite-sampling fluctuations from additional hardware
errors. Readout-error mitigation and temporal
symmetrization are applied to the experimental data
before Fourier reconstruction.

\begin{figure}[htbp]
\centering
\captionsetup[subfigure]{font=footnotesize,skip=2pt}


\begin{subfigure}[b]{0.47\linewidth}
    \centering
    \includegraphics[width=\linewidth]{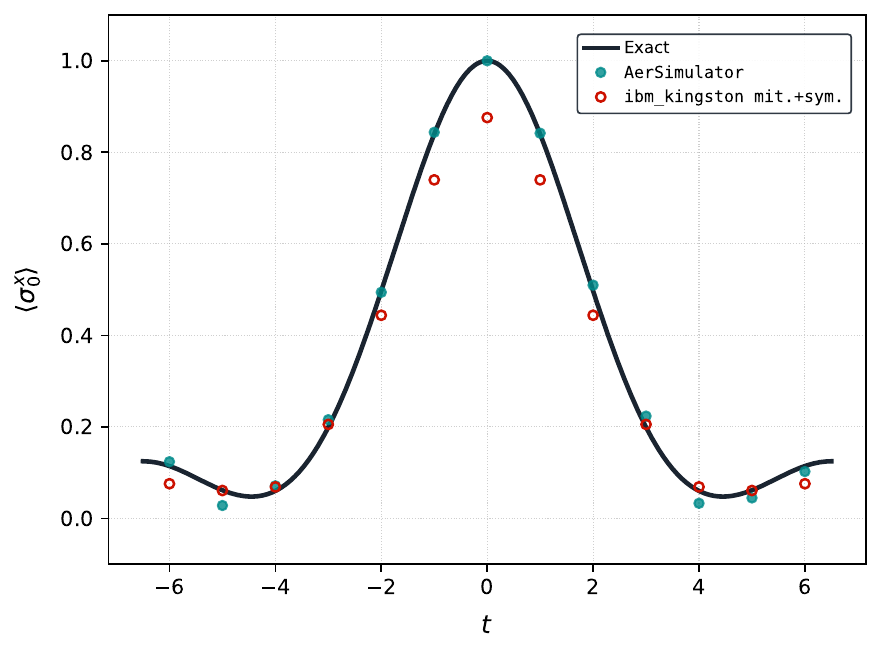}
    \caption{}
    \label{fig:sigma_x_12}
\end{subfigure}
\hfill
\begin{subfigure}[b]{0.47\linewidth}
    \centering
    \includegraphics[width=\linewidth]{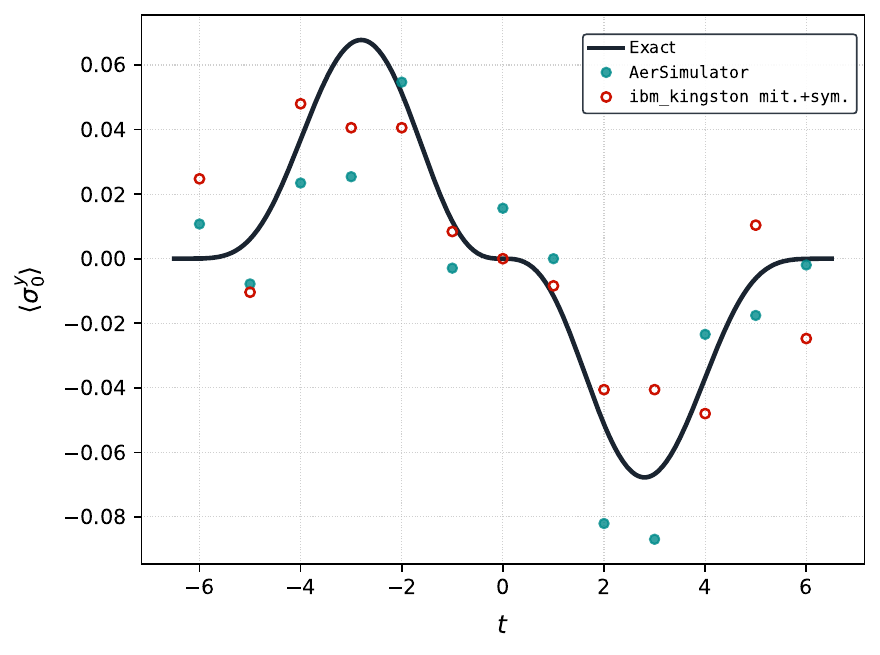}
    \caption{}
    \label{fig:sigma_y_12}
\end{subfigure}

\vspace{0.15cm}


\begin{subfigure}[b]{0.47\linewidth}
\centering

\begin{tikzpicture}
\begin{axis}[
    xlabel={$E/J$},
    ylabel={Degeneracy},
    xmin=-10, xmax=26,
    ymin=-20, ymax=1600,
    xtick={-8,-4,0,4,8,12,16,24},
    grid=both,
    grid style={line width=.1pt, draw=gray!15},
    major grid style={line width=.2pt, draw=gray!40},
    legend pos=north east,
    legend style={
        font=\tiny,
        nodes={inner sep=0.9pt},
        row sep=0.1ex
    },
    label style={font=\small},
    tick label style={font=\small},
    width=\linewidth,
    height=0.75\linewidth
]

\addplot[
    blue!40!black,
    mark=*,
    mark size=1.8pt,
    ycomb,
    line width=1.8pt
] coordinates {
(-8,450)
(-4,960)
(0,1432)
(4,768)
(8,396)
(12,64)
(16,24)
(24,2)
};
\addlegendentry{Exact}

\addplot[
    color={rgb,255:red,0; green,139; blue,139},
    mark=square*,
    mark size=1.7pt,
    ycomb,
    line width=0.8pt
] coordinates {
(-8,442)
(-4,964)
(0,1416)
(4,790)
(8,376)
(12,36)
(16,70)
(24,2)
};
\addlegendentry{\texttt{AerSimulator} post.}

\addplot[
    red,
    mark=o,
    mark size=1.7pt,
    ycomb,
    line width=0.8pt
] coordinates {
(-8,394)
(-4,972)
(0,1466)
(4,796)
(8,362)
(12,70)
(16,0)
(24,2)
};
\addlegendentry{\texttt{ibm\_kingston} post.}

\end{axis}
\end{tikzpicture}

\caption{}
\label{fig:spectral_constrained_12}
\end{subfigure}
\hfill
\begin{subfigure}[b]{0.47\linewidth}
    \centering
    \includegraphics[width=\linewidth]{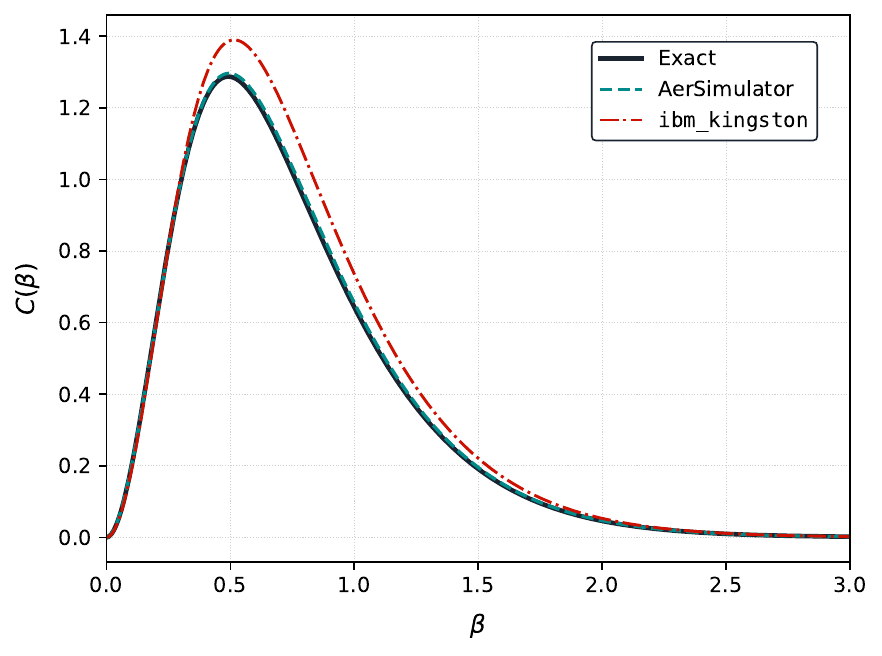}
    \caption{}
    \label{fig:heat_capacity_12}
\end{subfigure}

\caption{
Results for the 12-spin Kagome cluster.
(a,b) Ancilla expectation values
$\langle\sigma_0^x\rangle$ and $\langle\sigma_0^y\rangle$.
(c) Reconstructed energy-level degeneracies.
(d) Heat capacity $C(\beta)$.
Exact results are compared with shot-based
\texttt{AerSimulator} and experimental
\texttt{ibm\_kingston} data.
}
\label{fig:kagome_report_12}

\end{figure}
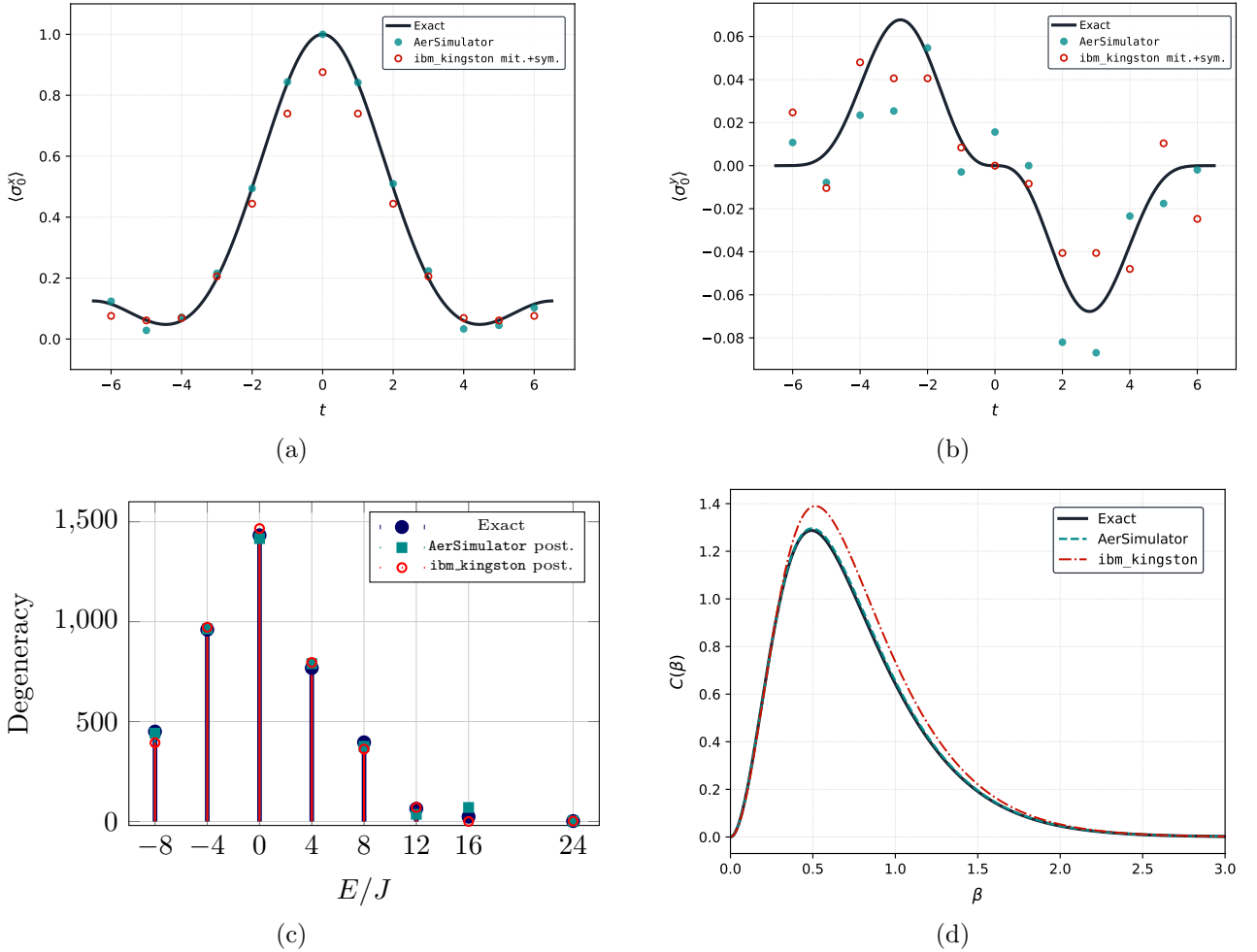

For this cluster, the Fourier grid contains $Q=13$
points, and Eq.~\eqref{eq:shot_noise_bound} gives
$
\sigma_{g_k} \simeq 25.1,
$
corresponding to a nominal $3\sigma$ threshold of
approximately $75$ states. Both datasets are processed
using the same reconstruction procedure as for $N=6$,
including noise-threshold filtering, normalization to
$\sum_k g_k=2^{12}$, and the physical constraints
on degeneracies. The results are presented in
Fig.~\ref{fig:kagome_report_12}.


The \texttt{AerSimulator} results reproduce the dominant
spectral weights with good accuracy, while the hardware
data show larger deviations due to additional device
noise. Highly degenerate levels remain well resolved
in both cases, whereas levels with small degeneracies
are more difficult to reconstruct. These differences
also appear in the heat capacity: the simulator closely
follows the exact curve, while the hardware results
exhibit larger deviations but retain its main features.
The comparison illustrates the additional limitations
introduced by quantum hardware beyond the statistical
uncertainty associated with finite sampling.


\subsection{Larger Kagome clusters and circuit-depth optimization}

We next examine periodic Kagome clusters with $N=24$ and $N=36$,
where the increasing circuit complexity makes hardware noise
particularly relevant. To reduce the circuit cost, we implement
each three-body Ising interaction using the ancilla-parity decomposition
\begin{equation}
\label{eq:parity_block}
U_{ij}(\theta)=P_{ij}R_z(\theta)P_{ij},\qquad
P_{ij}=\mathrm{CX}_{i,a}\mathrm{CX}_{j,a},
\end{equation}
where $a$ denotes the ancilla. Since all interaction terms commute,
the edges can be reordered without introducing Trotter error.
The periodic Kagome graph has even vertex degrees, so its edges
admit an Euler circuit. For consecutive edges $(i,j)$ and $(j,k)$,
the common parity operations cancel,
\begin{equation}
\label{eq:euler_cancellation}
P_{ij}P_{jk}
=\mathrm{CX}_{i,a}\mathrm{CX}_{j,a}
\mathrm{CX}_{j,a}\mathrm{CX}_{k,a}
=\mathrm{CX}_{i,a}\mathrm{CX}_{k,a}.
\end{equation}
Thus, for $L=2N$ interactions, the logical two-qubit gate count
decreases from $4L$ to $2L+2$, i.e., from $192$ to $98$
for $N=24$ and from $288$ to $146$ for $N=36$.

After transpilation for \texttt{ibm\_kingston}, the $N=24$ circuits
have depths of $486$ and $488$ for the ancilla $X$ and $Y$
measurements, respectively, with $344$ CZ gates and a two-qubit
depth of $212$ in each case. For $N=36$, the corresponding depths
are $757$ and $759$, with $710$ CZ gates and a two-qubit depth
of $341$. The higher native-gate counts reflect the additional
routing and decomposition required by the hardware architecture.
Despite the logical optimization, these deep circuits remain
susceptible to accumulated gate errors and decoherence.

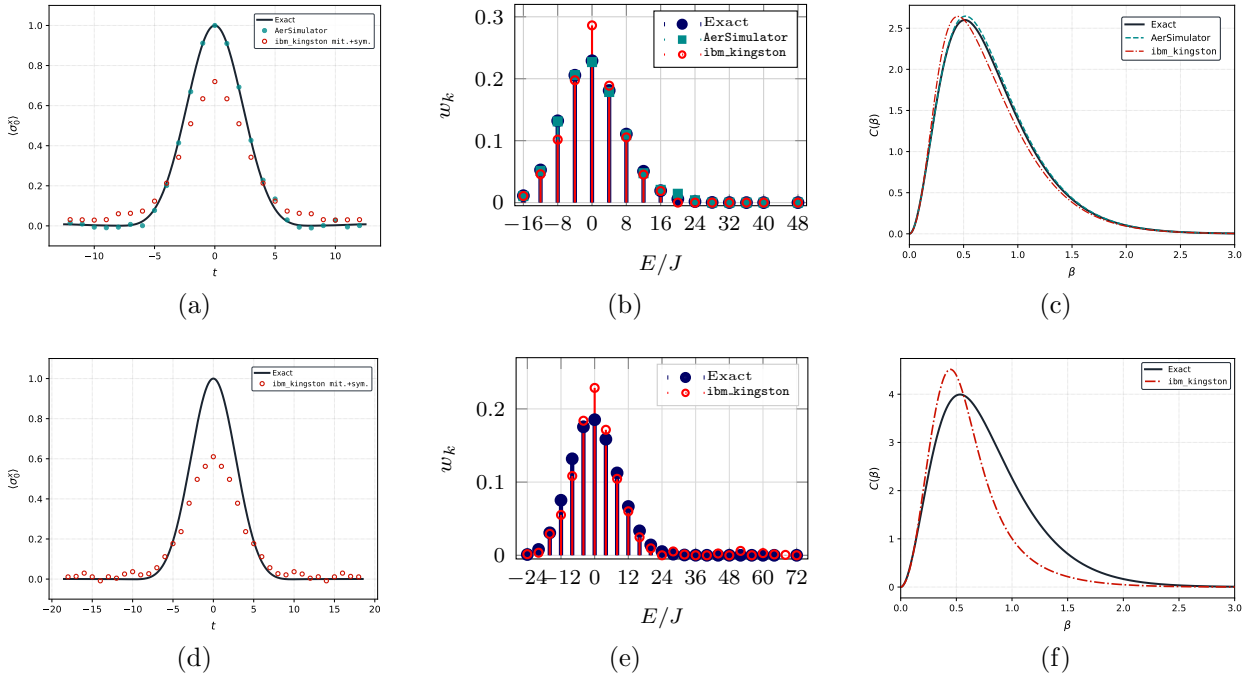
\begin{figure*}[htbp]
\centering

\captionsetup[subfigure]{
    font=footnotesize,
    skip=2pt,
    labelsep=none
}


\makebox[\textwidth][c]{%
\begin{subfigure}[t]{0.30\textwidth}
    \centering
    \includegraphics[width=\linewidth]{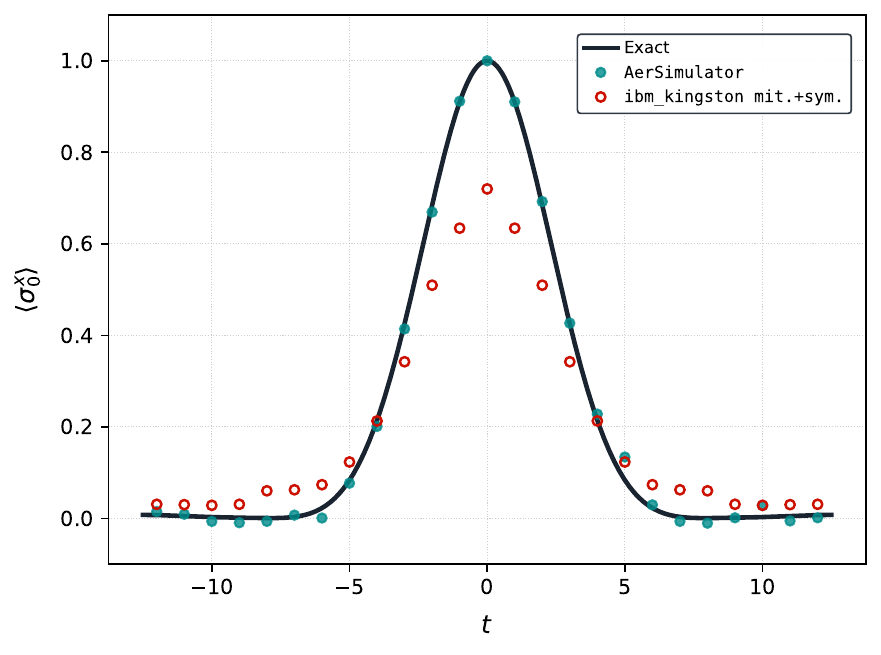}
    \caption{}
    \label{fig:sigma_x_24}
\end{subfigure}%
\hspace{0.025\textwidth}%
\begin{subfigure}[t]{0.32\textwidth}
\centering

\begin{tikzpicture}
\begin{axis}[
    xlabel={$E/J$},
    ylabel={$w_k$},
    xmin=-18, xmax=50,
    ymin=-0.005, ymax=0.32,
    xtick={-16,-8,0,8,16,24,32,40,48},
    grid=both,
    grid style={line width=.1pt, draw=gray!15},
    major grid style={line width=.2pt, draw=gray!40},
    legend pos=north east,
    legend style={
        font=\fontsize{5.2}{5.7}\selectfont,
        nodes={inner sep=0.4pt},
        row sep=0ex,
        cells={anchor=west}
    },
    legend image post style={scale=0.8},
    label style={font=\scriptsize},
    tick label style={font=\scriptsize},
    width=\linewidth,
    height=0.78\linewidth
]

\addplot[
    blue!40!black,
    mark=*,
    mark size=1.6pt,
    ycomb,
    line width=1.6pt
] coordinates {
(-16,0.011188626)
(-12,0.052742004)
(-8,0.132090569)
(-4,0.205642700)
(0,0.228995800)
(4,0.181022644)
(8,0.110630989)
(12,0.050628662)
(16,0.019261122)
(20,0.005729675)
(24,0.001645088)
(28,0.000320435)
(32,0.000091076)
(36,0.000007629)
(40,0.000002861)
(48,0.000000119)
};
\addlegendentry{Exact}

\addplot[
    color={rgb,255:red,0; green,139; blue,139},
    mark=square*,
    mark size=1.5pt,
    ycomb,
    line width=0.8pt
] coordinates {
(-16,0.010531187)
(-12,0.051216125)
(-8,0.131133080)
(-4,0.205905437)
(0,0.226919532)
(4,0.178623319)
(8,0.109608769)
(12,0.047434568)
(16,0.020582080)
(20,0.014272690)
(24,0.003773093)
(28,0)
(32,0)
(36,0)
(40,0)
(48,0.000000119)
};
\addlegendentry{\texttt{AerSimulator}}

\addplot[
    red,
    mark=o,
    mark size=1.5pt,
    ycomb,
    line width=0.8pt
] coordinates {
(-16,0.010678053)
(-12,0.045821071)
(-8,0.101743340)
(-4,0.197531939)
(0,0.286163568)
(4,0.188842058)
(8,0.105360031)
(12,0.045166373)
(16,0.018693447)
(20,0)
(24,0)
(28,0)
(32,0)
(36,0)
(40,0)
(48,0.000000119)
};
\addlegendentry{\texttt{ibm\_kingston}}

\end{axis}
\end{tikzpicture}

\caption{}
\label{fig:spectral_constrained_24}
\end{subfigure}%
\hspace{0.025\textwidth}%
\begin{subfigure}[t]{0.30\textwidth}
    \centering
    \includegraphics[width=\linewidth]{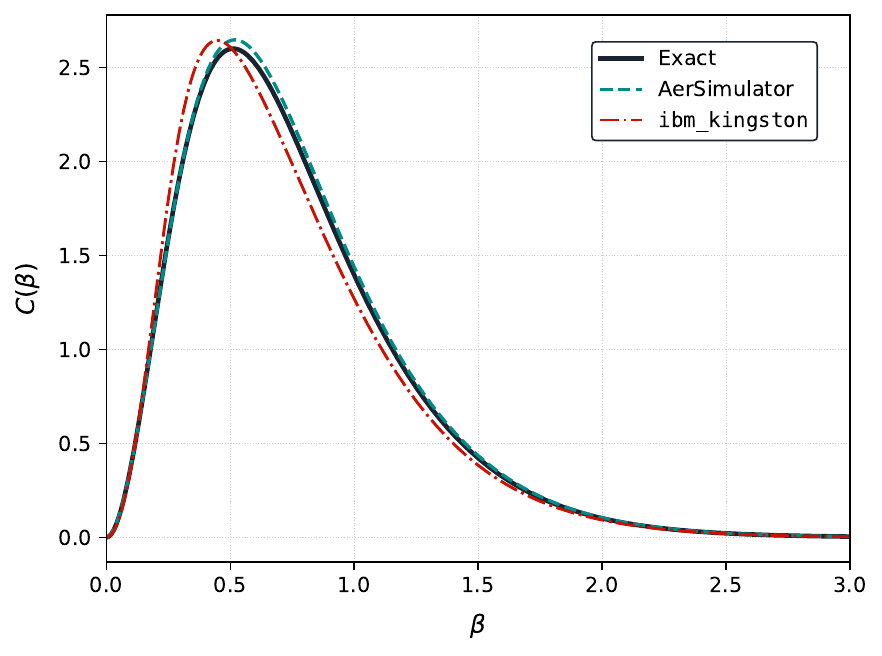}
    \caption{}
    \label{fig:heat_capacity_24}
\end{subfigure}%
}

\vspace{0.35cm}


\makebox[\textwidth][c]{%
\begin{subfigure}[t]{0.30\textwidth}
    \centering
    \includegraphics[width=\linewidth]{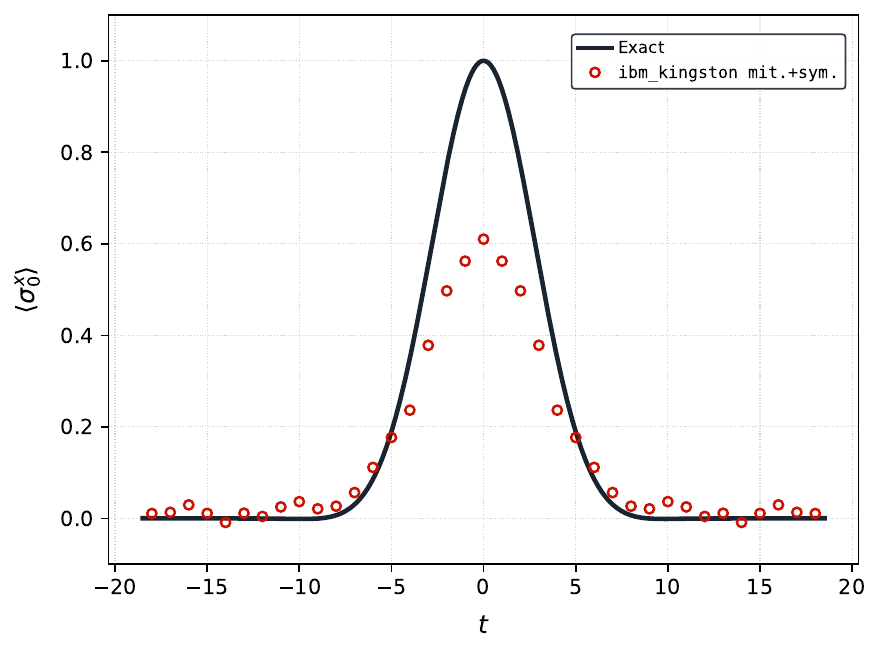}
    \caption{}
    \label{fig:sigma_x_36}
\end{subfigure}%
\hspace{0.025\textwidth}%
\begin{subfigure}[t]{0.32\textwidth}
\centering

\begin{tikzpicture}
\begin{axis}[
    xlabel={$E/J$},
    ylabel={$w_k$},
    xmin=-28, xmax=76,
    ymin=-0.005, ymax=0.27,
    xtick={-24,-12,0, 12, 24,36, 48, 60, 72},
    ytick={0,0.1,0.2},
    scaled ticks=false,
    tick label style={
        font=\scriptsize,
        /pgf/number format/fixed
    },
    grid=major,
    major grid style={
        line width=.2pt,
        draw=gray!30
    },
    legend pos=north east,
    legend style={
        font=\fontsize{5.2}{5.7}\selectfont,
        nodes={inner sep=0.4pt},
        row sep=0ex,
        cells={anchor=west},
        fill=white,
        fill opacity=0.9,
        text opacity=1,
        draw=gray!40
    },
    legend image post style={scale=0.8},
    label style={font=\scriptsize},
    width=\linewidth,
    height=0.78\linewidth
]

\addplot[
    blue!40!black,
    mark=*,
    mark size=1.6pt,
    ycomb,
    line width=1.6pt
] coordinates {
(-24,0.001034353)
(-20,0.007993600)
(-16,0.030716181)
(-12,0.075172631)
(-8,0.131817640)
(-4,0.175530276)
(0,0.185228602)
(4,0.158604451)
(8,0.112518277)
(12,0.066678835)
(16,0.033401277)
(20,0.014127810)
(24,0.005099486)
(28,0.001552826)
(32,0.000408581)
(36,0.000091799)
(40,0.000019034)
(44,0.000003512)
(48,0.000000681)
(52,0.000000117)
(56,0.000000026)
(60,0.000000003)
(64,0.000000001)
(72,0.0000000000291)
};
\addlegendentry{Exact}

\addplot[
    red,
    mark=o,
    mark size=1.5pt,
    ycomb,
    line width=0.8pt
] coordinates {
(-24,0.001545786)
(-20,0.003483291)
(-16,0.029270722)
(-12,0.055163316)
(-8,0.108351016)
(-4,0.183832908)
(0,0.228692986)
(4,0.171449330)
(8,0.104466612)
(12,0.060087319)
(16,0.024626484)
(20,0.009491462)
(24,0)
(28,0.005239008)
(32,0.001038487)
(36,0)
(40,0)
(44,0.002458541)
(48,0)
(52,0.005991608)
(56,0)
(60,0.002965026)
(64,0.001189095)
(68,0.000571666)
(72,0.0000000000291)
};
\addlegendentry{\texttt{ibm\_kingston}}

\end{axis}
\end{tikzpicture}

\caption{}
\label{fig:spectral_constrained_36}
\end{subfigure}%
\hspace{0.025\textwidth}%
\begin{subfigure}[t]{0.30\textwidth}
    \centering
    \includegraphics[width=\linewidth]{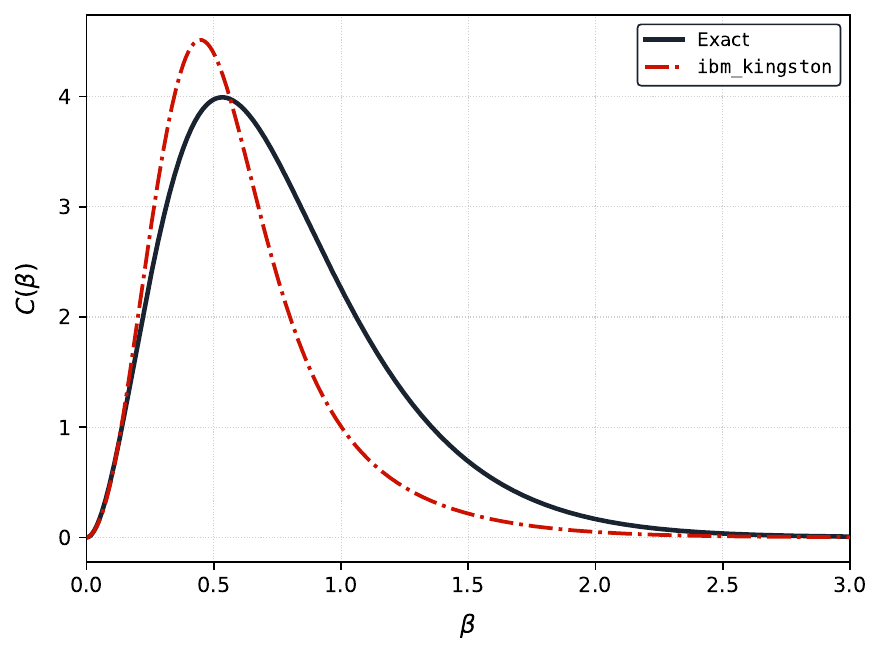}
    \caption{}
    \label{fig:heat_capacity_36}
\end{subfigure}%
}


\caption{
Reconstruction results for periodic Kagome clusters with
(a--c) $N=24$ and (d--f) $N=36$.
(a,d) Ancilla expectation values $\langle\sigma_0^x(t)\rangle$;
the $\langle\sigma_0^y(t)\rangle$ components are also used
in the reconstruction but are not shown.
(b,e) Post-processed normalized spectral weights compared
with exact results. Readout-mitigated \texttt{ibm\_kingston}
data are presented for both clusters, with additional
shot-based \texttt{AerSimulator} results for $N=24$.
(c,f) Heat capacities calculated from the corresponding
reconstructed spectra and compared with exact results.
}
\label{fig:kagome_large_clusters}

\end{figure*}

Both the $N=24$ and $N=36$ clusters are investigated
on the \texttt{ibm\_kingston} quantum processor, with
readout-error mitigation applied to the experimental data.
For $N=24$, we additionally perform noiseless, shot-based
\texttt{AerSimulator} calculations to distinguish
finite-sampling fluctuations from hardware-induced errors.
The measurements use $Q=25$ and $Q=37$ time points,
with $M=4096$ and $M=6400$ shots per setting for
$N=24$ and $N=36$, respectively.
The reconstructed spectra are compared with exact
classical results using the normalized spectral weights
$w_k=g_k/2^N$, as shown in
Fig.~\ref{fig:kagome_large_clusters}.

For $N=24$, the simulator closely reproduces the dominant
spectral weights, while the hardware results exhibit larger
deviations due to device noise. These deviations become more
pronounced for $N=36$, making weak spectral components
particularly difficult to resolve. The shot-noise bound
$\delta w_k\leq 1/\sqrt{QM}$ gives approximately
$3.1\times10^{-3}$ and $2.1\times10^{-3}$ for $N=24$
and $N=36$, respectively, whereas the exact ground-state
weight decreases from $1.1\times10^{-2}$ to about
$1.0\times10^{-3}$. For $N=36$, noise prevented reliable
reconstruction of the low-energy spectral edge. We therefore
used the known ground-state energy as an additional
constraint, without imposing its degeneracy.

For $N=36$, reducing this statistical upper bound below
the ground-state weight would require
$M>2.53\times10^4$ shots per measurement setting,
compared with the $6400$ used in our experiment.
Limited sampling precision and hardware noise may both
contribute to the deviations in the reconstructed heat
capacity. Increasing the number of shots would reduce
statistical fluctuations but would not remove systematic
hardware errors. Despite these limitations, the reconstructed
spectrum retains its dominant features, and the heat capacity
follows the overall temperature dependence of the exact
result. These results indicate that, with the stated
post-processing and ground-state energy constraint,
thermodynamic information can be recovered even from
the $N=36$ circuits containing $710$ CZ gates.

\section{Discussion}

The proposed method is closely related to quantum spectral
estimation based on real-time measurements of
$\mathrm{Tr}\,e^{-iHt}$~\cite{ferris2023,vangoffrier2026}.
While these approaches address general Hamiltonians through
finite-time Fourier analysis, our method exploits the known
discrete structure of the energy spectrum. The inverse DFT
provides exact reconstruction of the degeneracies in the
absence of noise, without spectral broadening or leakage.
For the periodic Kagome clusters considered here, the number
of sampling points is $Q=N+1$, growing linearly with system
size despite the exponential increase of the Hilbert-space
dimension. However, resolving small spectral weights may
require many shots, so this does not imply polynomial scaling
of the total measurement resources.

The circuit optimization can also be extended beyond graphs
admitting an Euler trail. Since the Ising interactions
commute, the edges can be partitioned into $r$ edge-disjoint
trails, with consecutive edges sharing a common vertex.
The parity operations can then be simplified along each
trail, reducing the logical two-qubit gate count from
$4L$ to $2L+2r$, where $L$ is the number of edges.
Thus, the optimization is applicable to general Ising
interaction graphs, although the reduction achieved
after transpilation depends on hardware connectivity.

The reconstruction accuracy is limited by statistical
fluctuations and hardware errors. While increasing the
number of shots reduces statistical uncertainty, systematic
errors require improved circuits and noise mitigation.
Readout mitigation, temporal symmetrization, and physical
constraints help stabilize the reconstruction but also
introduce correlations and nonlinear effects that are
not captured by simple shot-noise error bars. We therefore
provide analytical statistical bounds and assess hardware
effects through comparisons with exact results and
noiseless simulations. A complete uncertainty analysis
would require propagating measurement errors through
the entire post-processing procedure.

An important observation is that thermodynamic quantities
can remain reasonably accurate even when individual
degeneracies contain noticeable errors. Since these
quantities depend on weighted spectral contributions,
errors at different energy levels may partially compensate.
This explains why the reconstructed heat capacities retain
their main features despite substantial circuit depth and
noise. However, low-temperature properties remain
particularly sensitive to errors in low-energy degeneracies.

The method can also be extended to quantum many-body
systems with spectra lying on a known discrete grid.
In this case, preparing the system in a maximally mixed
state allows the ancilla protocol to measure the normalized
trace of the evolution operator. Examples include
stabilizer Hamiltonians, such as the toric-code
model~\cite{kitaev2003}, and Heisenberg models with
uniform all-to-all interactions. The inverse DFT remains
applicable regardless of the structure of the eigenstates,
although controlled evolution may be more demanding
for noncommuting Hamiltonians.

Another promising direction involves systems with known
energy grids but without exact analytical solutions,
such as the three-dimensional Ising model, frustrated
Ising systems in commensurate magnetic fields, and
antiferromagnetic models on general graphs. In these
cases, the possible energies are known, while their
degeneracies contain nontrivial many-body information.
The present results demonstrate the feasibility of
accessing such information on noisy quantum hardware
and motivate further applications to more complex
spin systems.

\section*{Author Contributions}

Both authors contributed equally to this work. ChatGPT (OpenAI) was used for English language editing, grammatical corrections, and stylistic refinement of the manuscript. Both authors reviewed and approved the final manuscript and take full responsibility for its content and integrity.

\end{document}